\documentclass[fleqn,twoside]{article}
\usepackage{espcrc2,html}

\usepackage{graphicx}
\newcommand{\AmS}{{\protect\the\textfont2
  A\kern-.1667em\lower.5ex\hbox{M}\kern-.125emS}}

\newcommand{\nit}{N_{\mbox{\tiny it}}}
\newcommand{\mpr}{{\frac{m_{\mbox{\tiny PS}}}{m_{\mbox{\tiny V}}}}}

\usepackage[figuresright]{rotating}

\title{Cost of QCD simulations with $n_f=2$ dynamical Wilson fermions}

\author{Th.\ Lippert\\[6pt]
SESAM/T$\chi$L collaboration\\[6pt]
Department of Physics, University of Wuppertal, Germany}

\begin{document}

\begin{abstract}
  Cost estimates for simulations of full QCD with $n_f=2$ Wilson
  fermions by hybrid Monte Carlo are presented. The extrapolations are
  based on the average number of iterations, $\nit$, of the iterative
  solver within the fermionic part of the HMC molecular dynamics,
  which is closely related to the minimal eigenvalue of
  $M^{\dagger}M$. The cost formula is determined as a product of the
  scaling functions of iterative solver and integrated autocorrelation
  time of $1/\nit$ as function of the inverse lattice pseudoscalar mass.
  Timings by SESAM/T$\chi$L allow to fix the pre-factor. It is
  demonstrated that a 2-flavor dynamical determination of light hadron
  masses with a statistical precision comparable to the corresponding
  quenched results from CP-PACS is the appropriate task for a 100
  Tflops system. 
\end{abstract}

\maketitle

\section{INTRODUCTION}

Realistic QCD simulations with dynamical fermions require to operate
beyond the $\rho$ decay threshold and closer to the continuum limit.
High energy physics can profit from lattice QCD as soon as
systematical and statistical errors are comparable to or smaller as
experimental ones \cite{Ecfa:2000rp}.

In order to make best usage of the next generation QCD machines concerning
these physical goals, we should employ the most
appropriate lattice actions and simulation algorithms. Hence, cost
predictions are needed for different lattice discretizations and
algorithms beyond $\mpr=0.5$.

As suggested by R. Kenway (Edinburgh)\footnote{The Wuppertal and
  Edinburgh groups are members of the working group ``Algorithms''
  supported by the EU network HPRN-CT-2000-00145 Hadrons Lattice
  QCD.}, chairman of the Lattice 2001 panel discussion, we determine
the costs of the SESAM and T$\chi$L experiments to estimate the costs
of future hybrid Monte Carlo simulations with two degenerate flavors
of standard Wilson fermions.

%The predictions are presented in form of a simple scaling formula as
%given in \cite{ECFA:1999ea} to be able to compare with results from
%other groups using different actions.

\section{SESAM/T$\chi$L SIMULATIONS}

SESAM/T$\chi$L has generated 10 ensembles of full QCD vacuum
configurations with $O(5000)$ HMC trajectories each, at $\beta=5.6$
and $5.5$ in the region $0.57 < \mpr<0.85$ with two flavors of Wilson
fermions.  The lattice sizes are $16^3\times 32$ (SESAM) and
$24^3\times 40$ (T$\chi$L), corresponding to physical sizes (from
$\rho$ mass) of $1.372(36)$ fm (SESAM) and of $1.902(34)$ fm
(T$\chi$L) after chiral extrapolation.  We used the standard hybrid
Monte Carlo algorithm \cite{Duane:1987de}, boosted by BiCGStab as
solver \cite{Vorst:1992vd}, {\em ll}\,-SSOR preconditioning
\cite{Fischer:1996th} and the educated guess procedure (chronological
inversion method) \cite{Brower:1997vx}. Running primarily on APE100
systems at DESY/Zeuthen, DFG/Bielefeld and INFN/Rome, the total costs
of the simulations amount to about 0.06 Tflops-yrs.

In table \ref{table:1} we give important quantities from
SESAM/T$\chi$L. We have exploited both o/e and SSOR
preconditioned fermion actions \cite{Fischer:1996th}. The latter
series are used for the cost analysis, as SSOR 
shows better scaling behavior.

\begin{table*}[!htb]
\caption{Characteristic quantities from SESAM/T$\chi$L. The {\em ll}\,-SSOR trajectories
are indicated by \fbox{\small\#}.}
\label{table:1}
\renewcommand{\tabcolsep}{1.1pc} % enlarge column spacing
\renewcommand{\arraystretch}{.96} % enlarge line spacing
\begin{tabular*}{\textwidth}{|@{}@{\extracolsep{\fill}}l|c|cccc|cc|}
\hline
{$\beta$}  & & \multicolumn{4}{c|}{$16^3\times 32$} &  \multicolumn{2}{c|}{$24^3\times 40$}\\
\hline
$5.6$            &$\mpr$   & 0.83  & 0.81   & 0.76   & 0.68   & 0.70   & 0.57  \\
     &$T_{\rm equi}$ & {5200} & \fbox{5400} & {3220}/\fbox{2030} &{2350}/\fbox{2600} &  \fbox{4700} &  \fbox{4000} \\
     &$\tau_{\rm int}^{\nit}$ &  & 19(4) &  25(6) & 33(4) & 
     36(4) & 50(5) \\
\hline
$5.5$            &$\mpr$   & 0.85  & 0.80   & 0.75   & 0.68   & 
\multicolumn{2}{c|}{}\\
     &$ T_{\rm equi}$ & \fbox{3500} & \fbox{4000} & \fbox{5000}  & \fbox{5000} &
\multicolumn{2}{c|}{}\\
     &$\tau_{\rm int}^{\nit}$ & 19(2) & 24(3) & 38(2)
     & {47(3)} &  \multicolumn{2}{c|}{}\\
\hline
\end{tabular*}
\vspace*{-6pt}
\end{table*}

\section{SCALING FITS}

\begin{figure}[!htb]
\centerline{\includegraphics[width=.7\columnwidth]{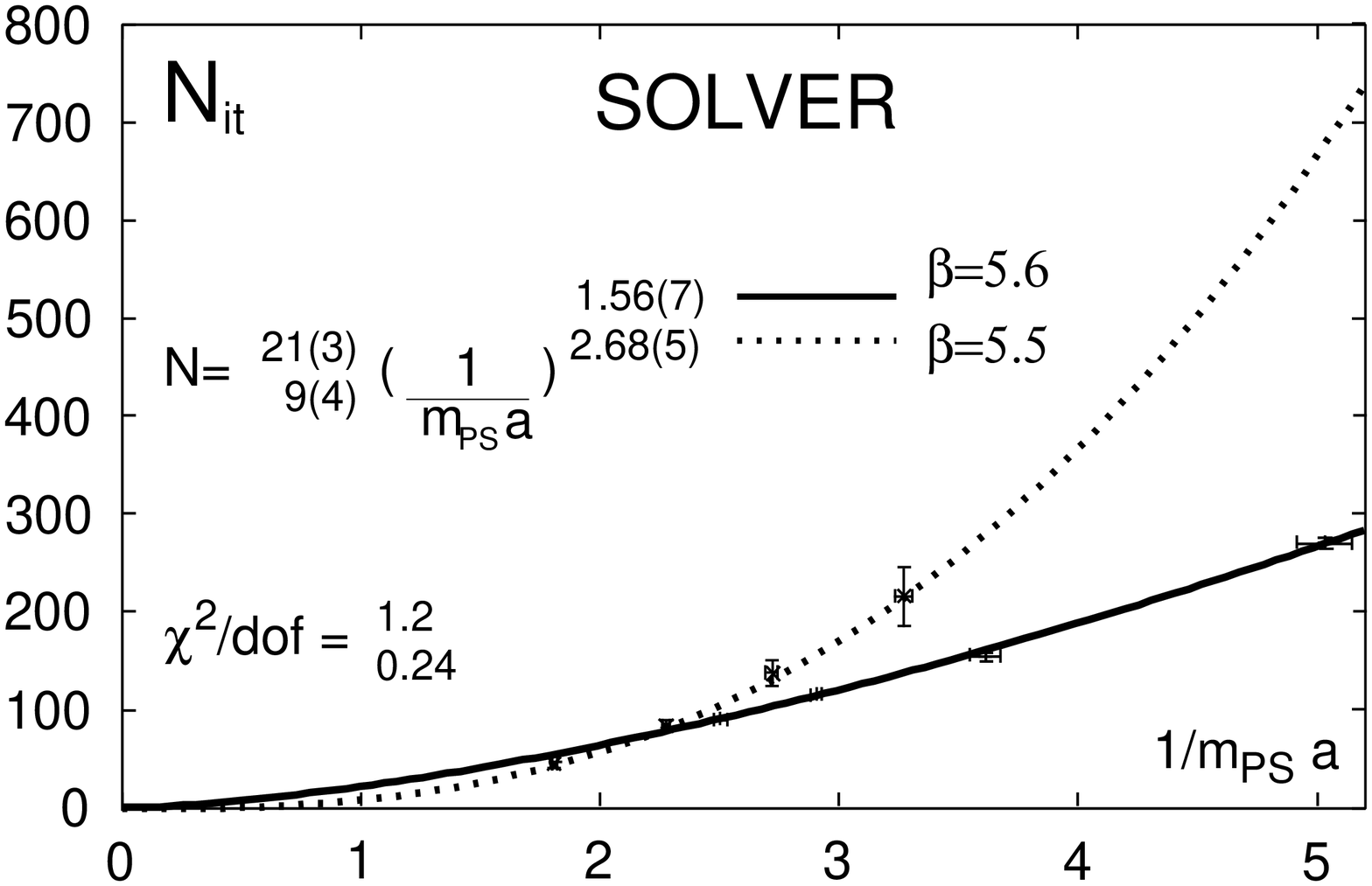}}
\centerline{\includegraphics[width=.7\columnwidth]{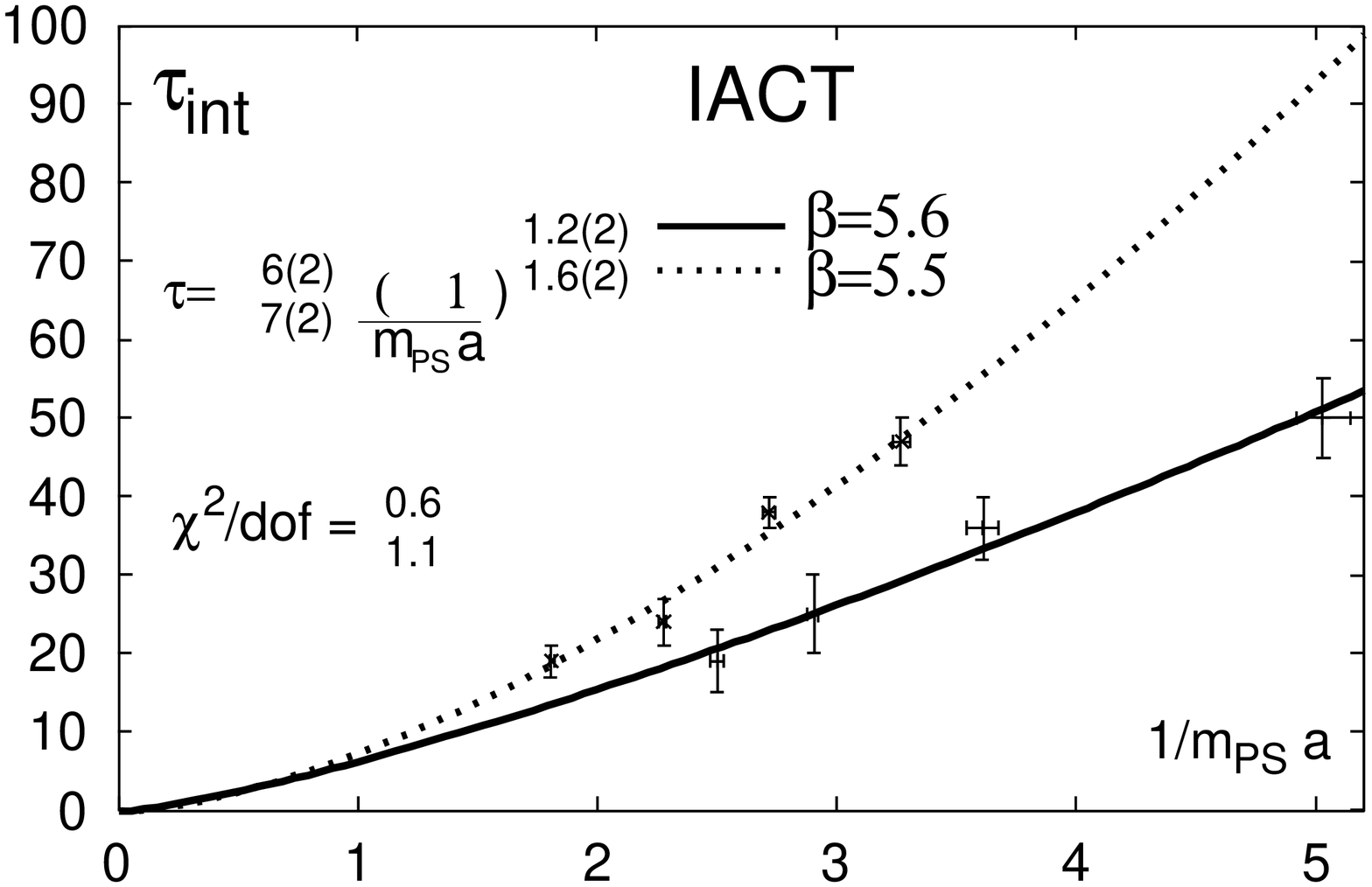}}
\vspace*{-10pt}
\caption{Scaling of iterative solver and integrated autocorrelation time.}
\label{SOLVER}
\vspace*{-10pt}
\end{figure}

Figure \ref{SOLVER}a presents fits to the average number of
iterations, $\nit$, of the {\em ll}\,-SSOR preconditioned BiCGStab
solver, as function of $1/m_{\mbox{\tiny PS}}a$.  Note
that the $\beta=5.6$ result scales much better than
$\beta=5.5$. Presumably, smoother gauge fields allow for better {\em
  ll}-SSOR preconditioning.

Figure \ref{SOLVER}b shows fits to the integrated autocorrelation
times of the series of $1/\nit$, the inverse number of iterations.
$1/\nit$ is related to the minimal eigenvalue of $M^{\dagger}M$. Its
autocorrelation time is comparable to that of the topological charge.
Again, we observe a strong $\beta$ dependence.

\begin{table*}[!htb]
\caption{Costs of simulations with 2 flavors of  Wilson fermions
in analogy to the quenched setting of CP-PACS.}
\label{table:2}
\renewcommand{\tabcolsep}{.6pc} % enlarge column spacing
\renewcommand{\arraystretch}{1.} % enlarge line spacing
\begin{tabular*}{\textwidth}{|@{}@{\extracolsep{\fill}}c|lllll|c|c|c|c|}
\hline
{$a[fm]\backslash\mpr$}  & 0.75 & 0.70 & 0.60 & 0.50 & 0.40 & $\sum$
[Tflops-yrs]& $z_2$ & $L/a$ & \# of confs\\
\hline
$0.102$   & 0.46 & 0.79 & 0.21 & 0.58 & 18 & 25(8)   & 4.3  & 32 & 800 \\
$0.076$   & 1.2  & 1.8  & 3.5  & 7.0  & 15 & 28(9)   & 2.8  & 40 & 600\\
$0.064$   & 3.7  & 6.7  & 12   & 23   & 50 & 95(35)  & 2.8 (!)  & 48 & 400 \\
$0.050$   & 17  & 28    & 60   & 120  & 260& 485(150)& 2.8 (!)  & 64 & 200 \\
\hline
\end{tabular*}
\vspace*{-6pt}
\end{table*}

\section{COST FORMULA}

Comparing the sustained CPU time for one HMC trajectory on APE100 with
the costs for the iterative solver allows to determine the
normalization of the cost function. Multiplication with the
autocorrelation time gives the effort to generate one statistically
independent configuration at given $\beta$ as function of
$1/m_{\mbox{\tiny PS}}a$. We assume the volume to scale like $(L/a)^5$ (unlike $(L/a)^{4.55}$ of Ref.~\cite{Ecfa:2000rp}) and
the temporal lattice extent to be twice as large as the spatial
extent $L$.\vspace*{-4pt}
\newcommand{\mycdot}{\!\cdot\!}
\begin{eqnarray}
\beta=5.6\!: N_{\mbox{\tiny flops}}=2.3(7)\mycdot 10^7\mycdot\left(\frac{L}{a}\right)^5\!\!\mycdot\left(\frac{1}{a\,m_{\mbox{\tiny
        PS}}}\right)^{2.8(2)}\!\!\!\!\!\!\!\!\!\!\!\!\!\!,\nonumber\\
\beta=5.5\!: N_{\mbox{\tiny flops}}=1.6(4)\mycdot 10^7\mycdot\left(\frac{L}{a}\right)^5\!\!\mycdot\left(\frac{1}{a\,m_{\mbox{\tiny
        PS}}}\right)^{4.3(2)}\!\!\!\!\!\!\!\!\!\!\!\!\!\!.\label{COST}
\end{eqnarray}

\section{EXTRAPOLATIONS}
\vspace*{-3pt}
The CP-PACS quenched simulations \cite{Aoki:1999yr} achieved finite
$a$ results for light hadrons with errors $< 1\% $ and continuum
result with errors between $1$ and $3 \%$. Linearly extrapolating
(\ref{COST}) in $a$ and $\mpr$, we find the upper bounds to the CPU
time (see table \ref{table:2}) needed to carry out an analogous
simulation with $n_f=2$ Wilson fermions.

\end{document}